\documentclass[printer]{aa}

\usepackage{graphicx}
\usepackage{txfonts}
\usepackage{natbib}
\usepackage{hyperref}
\usepackage[dvipsnames]{xcolor} 
\hypersetup{colorlinks=true,
            linkcolor=blue,
            citecolor=blue}
            
\bibpunct{(}{)}{;}{a}{}{,} 
          
\begin{document} 

   
\title{The Gaia-ESO Survey:\\Chemical evolution of Mg and Al in the Milky Way with Machine--Learning\thanks{Based on observations made with the ESO/VLT, at Paranal Observatory, under program 188.B-3002 (The Gaia-ESO Public Spectroscopic Survey, PIs G. Gilmore and S. Randich). Also based on observations under programs 171.0237 and 073.0234}}

   \author{M.~Ambrosch\inst{1}
          \and
          G.~Guiglion\inst{2}
          \and
          \v{S}.~Mikolaitis\inst{1}
          \and
          C.~Chiappini\inst{2}
          \and
          G.~Tautvai\v{s}ien{\. e}\inst{1}
          \and
          S.~Nepal\inst{2, 3}
          \and
          G.~Gilmore\inst{4}
          \and
          S.~Randich\inst{5}
          \and
          T.~Bensby\inst{6}
          \and
          M.~Bergemann\inst{7, 8}
          \and
          L.~Morbidelli\inst{5}
          \and
          E.~Pancino\inst{5}
          \and
          G.~G.~Sacco\inst{5}
          \and
          R.~Smiljanic\inst{9}
          \and
          S.~Zaggia\inst{10}
          \and
          P.~Jofr\'e\inst{11}
          \and
          F.~M.~Jim\'enez-Esteban\inst{12}
          }

   \institute{Institute of Theoretical Physics and Astronomy, Vilnius University, Saul\.etekio av. 3, 10257 Vilnius, Lithuania \\
   \email{markus.ambrosch@ff.vu.lt}
   \and
   Leibniz-Institut f\"ur Astrophysik Potsdam (AIP), An der Sternwarte 16, 14482 Potsdam, Germany
   \and
   Institut f\"ur Physik und Astronomie, Universit\"at Potsdam, Karl-Liebknecht-Str. 24/25, 14476 Potsdam, Germany
   \and
   Institute of Astronomy, University of Cambridge, Madingley Road,
   Cambridge CB3 0HA, United Kingdom
   \and
   INAF – Osservatorio Astrofisico di Arcetri, Largo E. Fermi 5, 50125
   Florence, Italy
   \and
   Lund Observatory, Department of Astronomy and Theoretical physics, Box 43, 221 00 Lund, Sweden
   \and
   Max Planck Institute for Astronomy, K\"onigstuhl 17, 69117, Heidelberg, Germany
   \and
   Niels Bohr International Academy, Niels Bohr Institute, University of Copenhagen Blegdamsvej 17, DK-2100 Copenhagen, Denmark
   \and
   Nicolaus Copernicus Astronomical Center, Polish Academy of Sciences, ul. Bartycka 18, 00-716, Warsaw, Poland
   \and
   INAF - Padova Observatory, Vicolo dell'Osservatorio 5, 35122
   Padova, Italy
   \and
   N\'ucleo de Astronom\'ia, Universidad Diego Portales, Ej\'ercito 441, Santiago, Chile
   \and
   Departmento de Astrof\'isica, Centro de Astrobiolog\'ia (CSIC-INTA), ESAC Campus, Camino Bajo del Castillo s/n, E-28692 Villanueva de la Canada, Spain
   }

   \date{Received ...; accepted  ...}

 
  \abstract
   {To take full advantage of upcoming large-scale spectroscopic surveys, it will be necessary to parameterize millions of stellar spectra in an efficient way. Machine--learning methods, and especially convolutional neural networks, will be one of the main tools for achieving this task.}
   {We aim to prepare the machine--learning ground for the next generation of spectroscopic surveys, such as 4MOST and WEAVE. Our goal is to show that convolutional neural networks can predict accurate stellar labels from relevant spectral features in a physically meaningful way. The predicted labels can be used to investigate properties of the Milky Way galaxy.}
   {We built a neural network and trained it on GIRAFFE spectra with associated stellar labels from the sixth internal Gaia-ESO data release. Our network architecture contains several convolutional layers that allow the network to identify absorption features in the input spectra. Internal uncertainty is estimated from multiple network models. We used \textit{t-distributed Stochastic Neighbor Embedding} to remove bad spectra from our training sample.}
   {Our neural network predicts the atmospheric parameters \textit{T}\textsubscript{eff} and log(\textit{g}) as well as the chemical abundances [Mg/Fe], [Al/Fe], and [Fe/H] for 30\,115 stellar spectra. The internal uncertainty is 24~K for \textit{T}\textsubscript{eff}, 0.03 for log(\textit{g}), 0.02~dex for [Mg/Fe], 0.03~dex for [Al/Fe], and 0.02~dex for [Fe/H]. The network gradients reveal that the network is inferring the labels in a physically meaningful way from spectral features. We validated our methodology using benchmark stars and  recovered the properties of different stellar populations in the Milky Way galaxy.}
   {Such a study provides very good insights into the application of machine--learning for the analysis of large-scale spectroscopic surveys, such as WEAVE and 4MIDABLE-LR and -HR (4MOST Milky Way disk and bulge low- and high-resolution). The community will have to put a substantial effort in building proactive training sets for machine--learning methods to minimize the possible systematics.}


\titlerunning{Mg \& Al with Machine--Learning}
\authorrunning{Ambrosch et al.}

\maketitle
\clearpage
%
\section{Introduction}\label{Introduction}

The use of machine--learning for the exploration of big data sets in astronomy was predicted over three decades ago \citep{1988ESOC...28..245R}. Yet, the high computational costs of this method have long delayed its advance. Some of the first applications of neural networks, a sub-field of machine--learning, include the automatic detection of sources in astronomical images (SExtractor, \citealt{1996A&AS..117..393B}), the morphological classification of galaxies \citep{1996MNRAS.283..207L} and the classification of stellar spectra \citep{1997Obs...117..250B}. In recent years, the increasing power of modern computer systems and the possibilities of cloud computing have led to a growing popularity of machine--learning methods. Powerful open-source libraries such as TensorFlow \citep{tensorflow2015-whitepaper} and PyTorch \citep{NEURIPS2019_9015} for Python programming offer easy-to-use frameworks for building and training various types of neural networks. \par
Spectroscopic surveys provide insights into the evolution of individual stars, of large-scale structures such as globular clusters and of the Milky Way galaxy as a whole. Upcoming projects, for example the William Herschel Telescope Enhanced Area Velocity Explorer (WEAVE, \citealt{10.1117/12.2312031}) and the 4-metre Multi-Object Spectroscopic Telescope (4MOST, \citealt{2019Msngr.175....3D}) will observe millions of stars. Efficient automatic tools will be needed to analyze the large number of spectra that such surveys will deliver. \par
To determine the atmospheric parameters and chemical composition of stars, classical spectroscopic methods either measure equivalent widths of absorption lines or compare observed spectra to synthetic spectra. These synthetic spectra can be generated on-the-fly or are part of a pre-computed spectral grid. \cite{doi:10.1146/annurev-astro-091918-104509} provide an overview over classical spectral analysis methods in the context or large spectroscopic surveys. \par
Convolutional Neural-Networks (CNNs) have recently been used to simultaneously infer multiple stellar labels (i.e. atmospheric parameters and chemical abundances) from stellar spectra. Every CNN contains convolutional layers which enable the network to identify extended features in the input data. In stellar spectra these features are absorption lines and continuum points; In 2-D images such features could be eyes in a face or star clusters in a spiral galaxy \citep{2020AJ....160..264B}. Neural network methods are purely data-driven and therefore require no input of any physical laws or models. Instead, during a training phase the network learns to associate the strength of spectral features with the values of the stellar labels. This requires a training set of spectra with pre-determined labels, from which the network can learn. Training sets for spectral analysis typically contain several thousand stellar spectra with high quality labels. Current spectral surveys, which provide $\sim$10\textsuperscript{5} spectra with labels, are an ideal testing ground for the CNN approach to spectral parameterization. \par
Examples of stellar parameterization using CNNs can be found in several recent studies. \cite{2018MNRAS.475.2978F} have developed StarNet, a CNN that is able to infer the stellar atmospheric parameters directly from observed spectra in the APO Galactic Evolution Experiment (APOGEE, \citealt{2017AJ....154...94M}). A grid of synthetic spectra was used to train and test StarNet. 
Purely observational data from APOGEE DR14 were used by \cite{2019MNRAS.483.3255L} to train their astroNN convolutional network. To mimic the methods of standard spectroscopic analysis, astroNN is designed to use the whole spectrum when predicting atmospheric parameters but is limited to individual spectral features for the prediction of chemical abundances. \cite{2020A&A...644A.168G} trained their CNN on medium-resolution stellar spectra from the RAdial Velocity Experiment (RAVE, \citealt{Steinmetz_2020}) together with stellar labels that were derived from high-resolution APOGEE DR16 spectra. They also added absolute magnitudes and extinction corrections for their sample stars as inputs for the network. This information allowed their CNN to put additional constraints on its predictions of the effective temperature and surface gravity. \par
In this work, we propose to test a CNN approach in the context of the Gaia-ESO survey (GES, \citealt{2012Msngr.147...25G, 2013Msngr.154...47R}). We use GIRAFFE spectra with labels from the sixth internal data release. The GES survey is designed to complement the astrometric data from the Gaia space observatory \citep{2016A&A...595A...1G}. The goal 
of the present project is to prepare machine--learning ground for the next generation of spectroscopic surveys, such as 4MOST and WEAVE. This paper goes together with Nepal et al. (sub) that focuses on the chemical evolution of lithium with CNNs from GES GIRAFFE HR15N spectra.

This paper is organized as follows: In Sect.~\ref{Data} we present the data that we used to train and test our CNN. Section \ref{Network architecture and training} describes the architecture of our network and explains the details of the training process. The results of the training and the network predictions for the observed set are presented in Sect.~\ref{Training results}. In Sect.~\ref{Validation of results} we validate our results by investigating the CNN predictions for a number of benchmark stars. For the further validation we use our results to recover several properties of the Milky Way galaxy.

\section{Data}
\label{Data}

\subsection{Data preparation}
\label{Data preparation}

Our data set consists of spectra, associated stellar parameters, and abundances from the GES iDR6 data set. In the Gaia-ESO survey, atmospheric parameters and chemical abundances are determined by multiple nodes that apply different codes and methodologies to the same spectra. A summary of the determination of atmospheric parameters from the GIRAFFE spectra is given in \cite{2014A&A...567A...5R}. Further information about the determination of chemical abundances can be found in \cite{2014A&A...572A..33M}. The spectra were taken with the GIRAFFE spectrograph that covers the visible wavelength range of 370 - 900~nm. Several setups divide the whole GIRAFFE spectral range into smaller parts. For this study we chose the HR10 (533.9 - 561.9~nm, R = 19800) and HR21 (848.4 - 900.1~nm, R = 16200) setups because they cover important Mg and Al absorption features. \par
For our analysis we used normalized 1-D spectra from the GES archive. We removed bad pixels and cosmic ray spikes where necessary. To do so, we first calculated the median of all spectrum flux values. We then identified cosmic ray spikes by finding all pixels with flux values that exceeded this median flux by five sigma. The spikes were removed by setting their flux value to be equal to the spectrum median flux. Pixels with zero flux values were also set to the median flux. Afterwards, we corrected the spectra for redshift based on the radial velocity provided by GES. To reduce the number of pixels per spectrum and therefore the computational cost of the further analysis, we re-binned the spectra to larger wavelength intervals per pixel. The HR10 spectra were resampled to 0.06~$\AA$ per pixel and the HR21 spectra to 0.1~$\AA$ per pixel. After re-binning, the spectra were truncated at the ends to ensure that all spectra from one setup share the exactly same wavelength range. Eventually we combined the HR10 and HR21 spectra to create one input spectrum per star for our network. The combined spectra are composed of 8669 pixels each and cover the wavelength ranges from 5350-5600~$\AA$ and 8480-8930~$\AA$. \par
To build our training set, we performed several quality checks to ensure that our network will be trained on high-quality data. Spectra with signal-to-noise ratio (S/N) < 30 and large errors in atmospheric parameters and elemental abundances (\textit{eT}\textsubscript{eff} > 200~K, \textit{e}log(\textit{g}) > 0.3~dex, \textit{e}A(element) > 0.2~dex) were discarded, as well as spectra that were marked with the TECH or PECULI flags or have rotation velocities > 20~km\, s$^{-1}$. We also removed spectra that showed a difference larger than 0.2~dex between the provided metallicity [Fe/H] (as a stellar atmospheric parameter) and the \ion{Fe}{i} elemental abundance. \par
\begin{figure*}
\centering
  \includegraphics[width=0.9\textwidth]{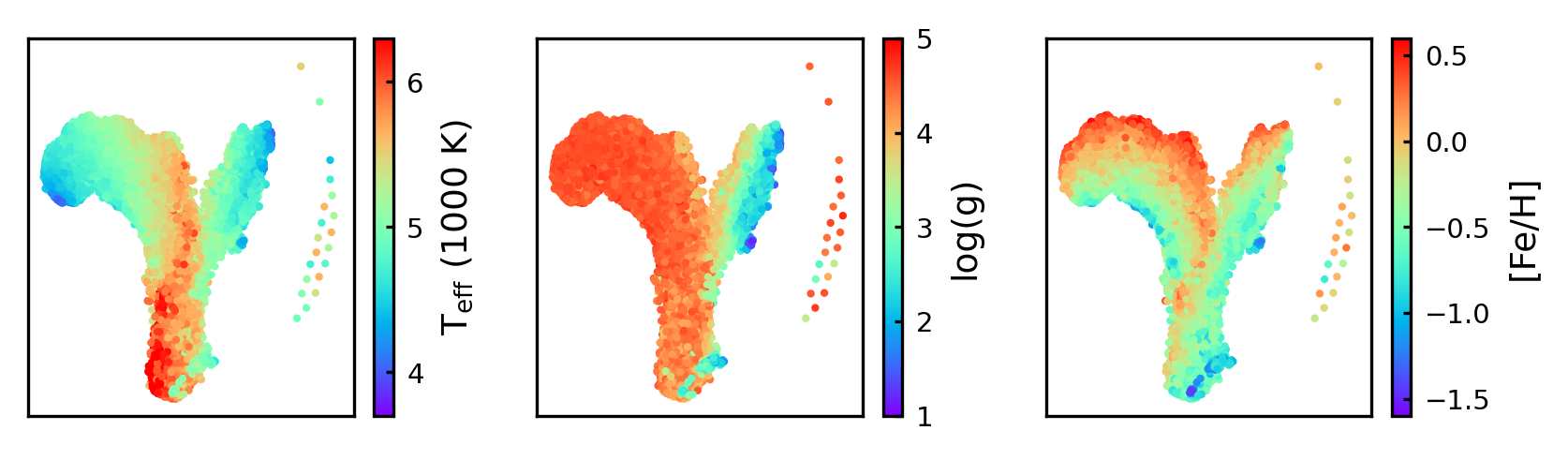}
  \caption{A t-SNE similarity map of our sample GIRAFFE spectra. The three panels show the same map, each color-coded with a different physical parameter. While the relative distance of points in the map indicate the degree of similarity of the corresponding spectra, their X and Y coordinates themselves have no physical meaning.}
  \label{fig:t-SNE}
\end{figure*}

We further examined the remaining spectra to find possible outliers and incorrect measurements. To investigate the similarity between all the spectra, a t-distributed stochastic neighbor embedding (t-SNE) analysis was employed. The t-SNE analysis is a popular technique to visualize the internal relationships and similarities in high dimensional data sets by giving each data point a location in a two- or three-dimensional similarity map \citep{JMLR:v9:vandermaaten08a}. In our case, the data points are the individual spectra and the data set is n-dimensional, where n is the number of pixels in each spectrum. Figure \ref{fig:t-SNE} shows a two-dimensional similarity map for our combined spectra, obtained with the \textit{sklearn.manifold} library for python programming \citep{JMLR:v12:pedregosa11a}. Every point in the map corresponds to one spectrum and the distance between the individual points is determined by the similarity of the shapes of the individual spectra. There are two main branches in the map with several sub-structures. The two branches represent spectra from stars in two distinct populations: Main sequence stars with surface gravity log(\textit{g}) $\gtrsim$ 3.5 and stars in the giant branch with lower log(\textit{g}) values. The different physical properties in stellar atmospheres are reflected in the shapes of their spectra which in turn determine their locations on the t-SNE map. This connection between physical parameters and spectral features is what our CNN learns during the training phase. We see several outlier-spectra in the map. Upon inspection, these spectra show signs of emission lines, have distorted absorption features or have suffered from failed cosmic removal or wrong normalisation. We excluded these outliers from the further analysis. For the analysis of future surveys such as WEAVE and 4MIDABLE-HR surveys, including emission line stars will be a necessity, as we expect many young stars to be observed.

Every training spectrum has a set of associated stellar labels. In our case these are the two atmospheric parameters \textit{T}\textsubscript{eff} and log(\textit{g}) and the chemical abundances [Mg/Fe], [Al/Fe], and [Fe/H]. In the GES iDR6 data set the elemental abundances are given as absolute abundance values A(Element). We calculated [Fe/H] and [Element/Fe] as follows: $\rm [Fe/H]=A(Fe)_{star}-A(Fe)_{\odot}$ and $\rm [Element/Fe]=A(Element)_{star}-A(Element)_{\odot}-[Fe/H]$. The absolute solar abundances were taken from \cite{2007SSRv..130..105G}, consistently with GES spectral analysis strategy. The decision to use these relative abundances instead absolute abundances for the training of our network is justified in Sect.~\ref{Learning from spectral features}. \par
Magnesium and aluminum abundances are known to be sensitive to non-local thermodynamical equilibrium (NLTE) effects (\citealt{2017ApJ...847...15B}, \citealt{2022arXiv220611070L}). These were were not considered by GES during the abundance analysis from GIRAFFE spectra. However, the NLTE effects are small for the Mg and Al absorption lines that are relevant for this study: $-$0.10~dex to 0.08~dex for Mg and 0.02~dex to 0.08~dex for Al, depending on the stellar atmospheric parameters. We therefore do not expect that NLTE will have a significant effect on the training of our network and the following scientific validation and did not attempt to correct the given GES abundances for NLTE effects.

After applying all of the constraints mentioned above, we were left with 14\,696 combined spectra with associated high-quality atmospheric parameters and elemental abundances. As explained in Sect.~\ref{Training set and test set}, these 14\,696 spectra will be randomly split into a training set and a test set for the training of our CNN.

\subsection{Parameter space of input labels}
\label{Parameter space of input labels}

\begin{figure}
  \includegraphics[width=0.9\columnwidth]{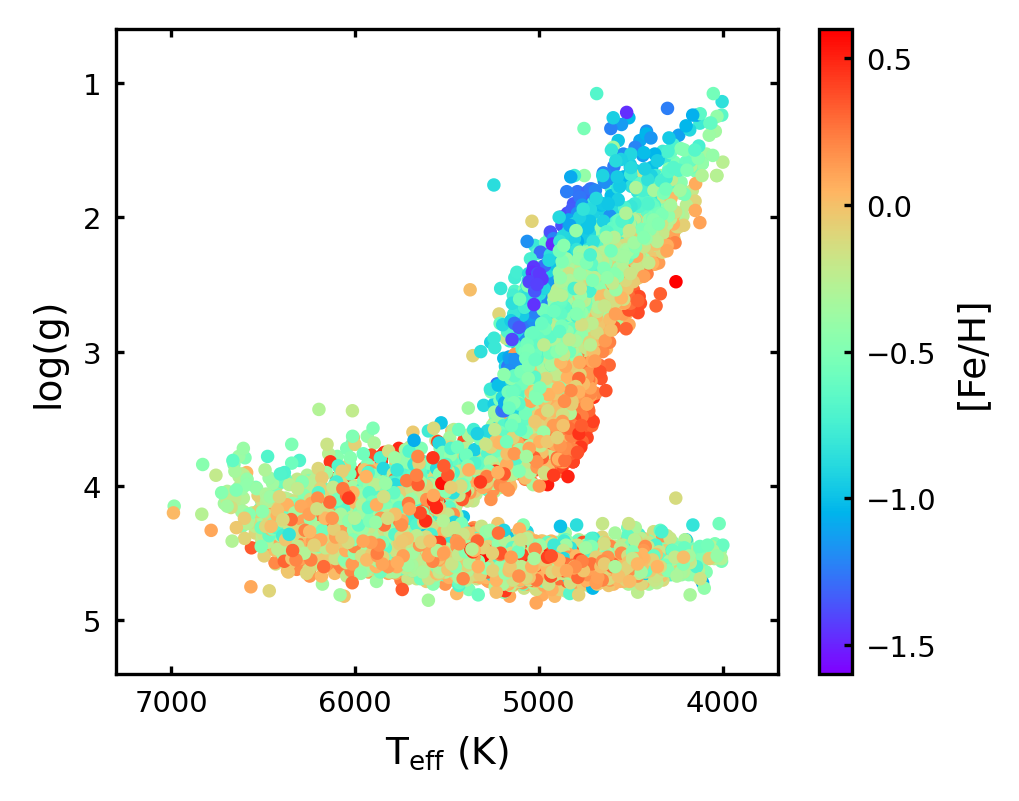}
  \caption{Kiel diagram containing the stars that will be used to train and test our neural network. The color-coding indicates the metallicity gradient in the giant branch stars.}
  \label{fig:kiel GES}
\end{figure}

\begin{figure}
  \includegraphics[width=0.9\columnwidth]{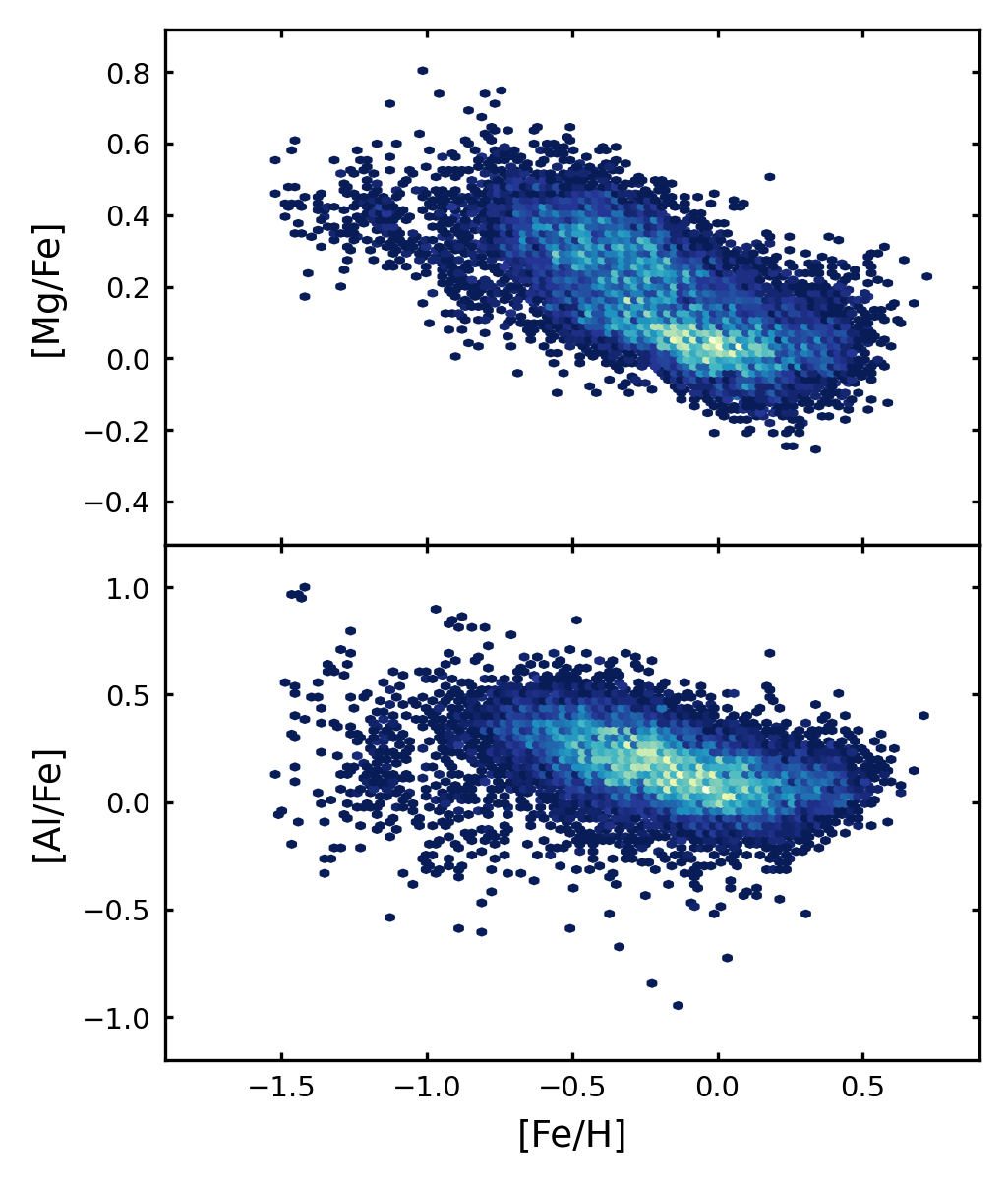}
  \caption{Density plots of [Mg/Fe] vs. [Fe/H] (top panel) and [Al/Fe] vs. [Fe/H] (bottom panel) for the 14\,696  stars in the training and test sets. Brighter colors indicate a higher density of data points.}
  \label{fig:MgAl_Fe_hist2d}
\end{figure}

To assess the parameter space of our training set input labels, we show the Kiel-diagram and abundance plots in Figs.~\ref{fig:kiel GES} and~\ref{fig:MgAl_Fe_hist2d}. Effective temperatures range from \textit{T}\textsubscript{eff} = 4000 - 6987~K, the surface gravity log(\textit{g}) is between 1.08 and 4.87~dex and [Fe/H] spans a range of $\sim$2~dex, from $-1.53$ to 0.72~dex. The color-coding in Fig.~\ref{fig:kiel GES} reveals the metallicity sequence in the giant-branch of the Kiel-diagram. \par Figure \ref{fig:MgAl_Fe_hist2d} shows density maps of the [Mg/Fe] and [Al/Fe] distribution of our training set. The [Mg/Fe] values range from $-0.25$ to 0.80~dex, [Al/Fe] values have a large spread of almost 2~dex, from $-0.95$ to 1.00~dex. The Mg distribution reveals two distinct regions of enhanced density, separated by a narrow region of lower density. These two regions reflect the separation of Milky Way stars into a thin-disk (low [Mg/Fe]) and a thick-disk (enhanced [Mg/Fe]) populations. Magnesium abundances are the best probe for this chemical separation between the thin- and thick-disk of our Galaxy (e.g. \cite{1998A&A...338..161F}, \citealt{2000A&A...358..671G}). As expected, we do not observe this separation in the [Al/Fe] plot. Our training set is dominated by nearby stars, due to the S/N cut and other quality criteria that we applied to the entire GES iDR6 data set. Therefore our data does not cover some of the Milky Way properties that become apparent when one investigates a larger volume of our galaxy. \cite{2020A&A...638A..76Q}, for example, find two detached [Al/Fe] sequences for stars close to the galactic center (R\textsubscript{Gal} < 2~kpc) in their sample of APOGEE stars. At low [Fe/H] several groups of stars can be observed in both the Mg and Al plots. The stars in these patches belong to different globular clusters. In the [Al/Fe] plot, the scatter of Al abundances in the globular clusters is considerably higher than the scatter of Mg at equal metallicities. This large spread of Al abundances, especially in globular clusters at low metallicities has already been observed in earlier GES releases (Fig.~4 in \citealt{2017A&A...601A.112P}) and indicates the existence of multiple stellar populations within the clusters.

\subsection{Observed set}

In addition to these spectra with high quality GES parameters, we composed set of spectra with S/N between 20 and 30. We call this the "observed set". The observed set will be used to test the performance of our CNN on spectra that were not involved in the training process. We did not put any quality constraints on the input labels for the spectra in the observed set, and a number of them do not have any reported Mg and Al abundances. As for the training and test sets, we removed spectra that were labeled with the TECH and PECULI flags and outliers in the t-SNE map. The ability of neural networks to extrapolate to labels outside of the parameter space of the training data is limited. Therefore we excluded spectra with GES labels outside of the training data distribution from our observed set. After applying these criteria to the GES iDR6 data set, our observed set contains 15\,419 spectra, most of them with associated GES labels.

\section{Network architecture and training}
\label{Network architecture and training}

A CNN acts as a function with many free parameters. In our case, this function takes stellar spectra as an input and outputs the associated atmospheric parameters and abundances. The network architecture then describes the shape of this neural network function. The goal of the training process is to find the optimal values of the free CNN parameters to accurately parameterize the input stellar spectra. In the following subsections we describe how a neural network can "learn" how to accurately parameterize stellar spectra. Our CNN was built and trained in a Python programming environment with the open source deep-learning library Keras \citep{chollet2015keras} using the TensorFlow back-end \citep{tensorflow2015-whitepaper}. 

\subsection{Network architecture}
\label{Network architecture}

\begin{figure}
  \centering
  \includegraphics[width=0.5\columnwidth]{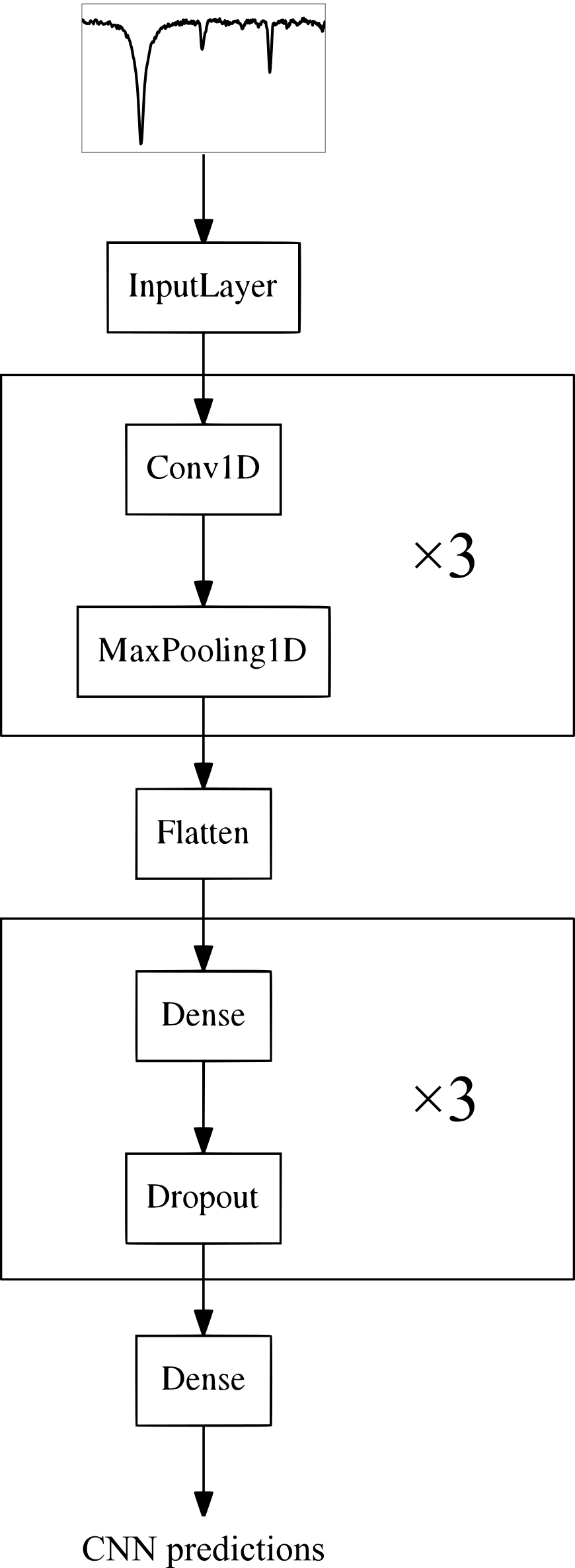}
  \caption{Architecture of our CNN. The input layer reads in the flux information of the stellar spectra. It is followed by three pairs of convolution and max-pooling layers. The filter outputs from the third convolution and max-pooling pair are then flattened to serve as inputs for the dense layers. Three dense layers (with a dropout layer after each) interpret the spectral features, found by the convolution layers, into output labels. The outputs from a last dense layer are the values of our six stellar labels (atmospheric parameters and elemental abundances).}
  \label{fig:model architecture}
\end{figure}

The different parts of a neural network architecture, the "layers", fulfil different purposes in the process of parameterizing stellar spectra. Our neural network consists of two main types of layers: Convolution layers that identify features and patterns in the input spectra, and dense layers which associate those spectral features to the output stellar parameters. A visualisation of our network architecture can be seen in Fig.~\ref{fig:model architecture}.

\subsubsection{Convolution layers}
\label{Convolution layers}

To identify the spectral features that correlate with the stellar labels, our CNN is composed of convolution layers. These layers convolve the input spectra with a number of 1-dimensional filters. The filters move across the input spectra and produce feature maps, which are the results of the spectrum-filter convolutions. While the length and number of filters is fixed, the purpose of each filer is learned during the training phase. The neural network learns how to adjust the filter values to achieve the best label predictions. Multiple convolution layers with multiple filters each can be put in sequence in a neural network architecture. Filters in one convolution layer then extract features in the feature maps that were produced by the previous convolution layer. Our CNN has three convolution layers with an increasing number of filters in each layer.

\subsubsection{Dense layers}
\label{Dense layers}

In order to build a high-dimensional complex function between the feature maps from the last convolution layer and the labels, so-called dense layers are necessary. Each dense layer consists of a fixed number of artificial neurons. An artificial neuron receives inputs from a previous layer, multiplies every input with its associated weight, and then passes the result to the neurons of the next dense layer. In this way, every neuron in one dense layer is connected to all neurons of the previous layer and to all neurons of the following layer (this is the reason why dense layers are also called "fully connected" layers). The last layer in a CNN is a dense layer where the number of neurons is equal to the number of labels that the network is designed to predict (in our case 5).

\subsubsection{Activation function}
\label{Activation function}

The relations between spectral features and physical stellar labels are non-linear. To reflect this non-linearity in our network training process, activation functions are used.
Activation functions transform the output of the convolution filters and the artificial neurons before they are passed on to the next layer. In recent machine--learning applications the "Leaky ReLU" activation function is most frequently used. It leaves positive and zero output values unchanged and multiplies negative outputs with a small positive value. Or, notated mathematically \citep{Maas2013RectifierNI}:

\
    \\
    $
    f(x)=
    \begin{cases}
    a \cdot x   &\text{if $x<0$} \\
    x            &\text{otherwise,}
    \end{cases}
    $
    \\
\

where $x$ is a filter or neuron output value before it is passed to the next layer. For our network, we adopt a Leaky ReLU activation function with $a = 0.3$ for all layers.

\subsubsection{Max-pooling and dropout}
\label{Max-pooling and dropout}

Over-fitting occurs when the network is very accurate in predicting the labels of the training set but shows a poor performance when predicting labels for the test set or an external observed set. In this case the network is not generalizing well for inputs outside of the training data. This is often the case when the network architecture is complex and the number of weights and biases is too large. In this context, max-pooling and dropout are popular regularisation devices used to prevent over-fitting during the training of a CNN.\par
Max-pooling helps to prevent over-fitting by reducing the complexity of the feature maps that are produced by the convolution layers. This is achieved by keeping only the highest value within a defined interval in every feature map. In this way the less important pixels of a feature map are discarded and the network is able to focus on pixels that show a strong response to the convolution filters. \par
Applying dropout after a dense layer randomly deactivates the output of a fraction of the layer neurons (these neurons are "dropped"). The weights associated with dropped neurons are therefore not updated for one training epoch (one passage of the entire training set through the network, see Sect. \ref{Epochs and batches}). After every epoch all neurons are reactivated and a new collection of neurons is dropped for the next epoch. As a consequence, the network architecture changes slightly after every epoch during the training. This prevents the network from relying too much on individual parts of the architecture and therefore individual features in the input spectrum. In this way the network is forced to learn from the whole spectrum which leads to a good generalization for different input spectra.

\subsection{Network training}
\label{Network training}

\begin{figure*}
 \centering
  \includegraphics[width=0.9\textwidth]{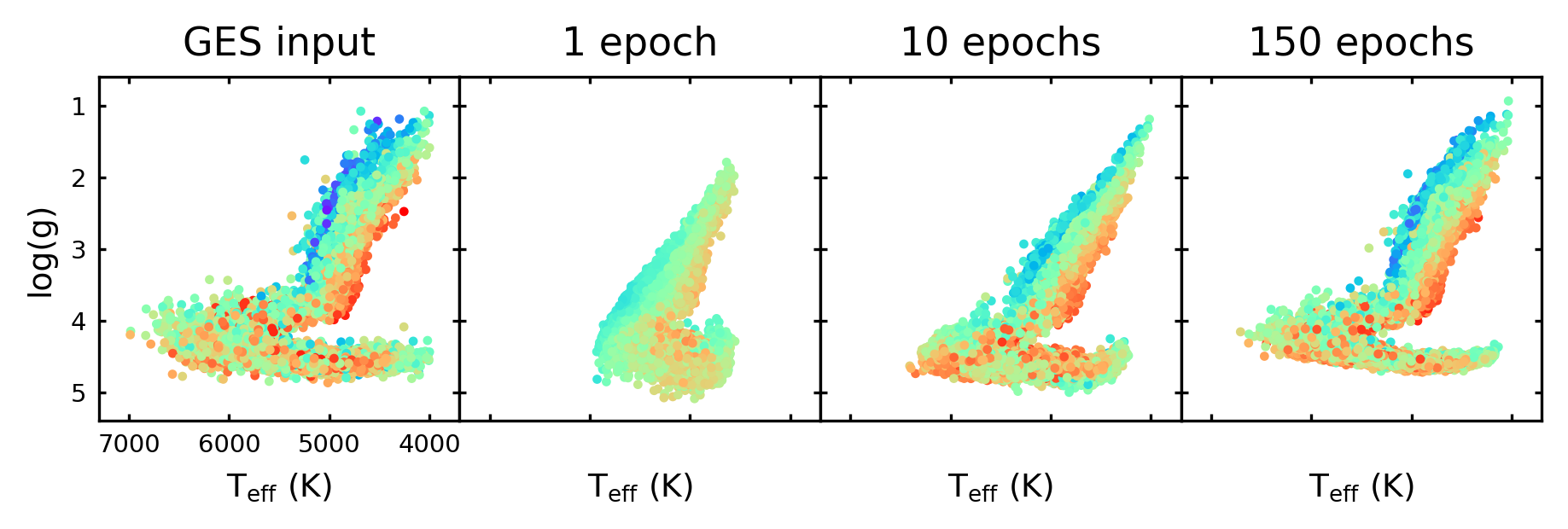}
  \caption{Evolution of the prediction-based Kiel-diagram during the network training. The far-left panel shows the Kiel-diagram based on the GES input values of \textit{T}\textsubscript{eff} and log(\textit{g}). Succeeding panels show the Kiel-diagram based on network predictions after 1, 10 and 150 training epochs. The color-coding, indicating the [Fe/H] values of each data point, is on the same scale as in Fig.~\ref{fig:kiel GES}.}
  \label{fig:kiel training}
\end{figure*}

\begin{figure}
  \includegraphics[width=0.9\columnwidth]{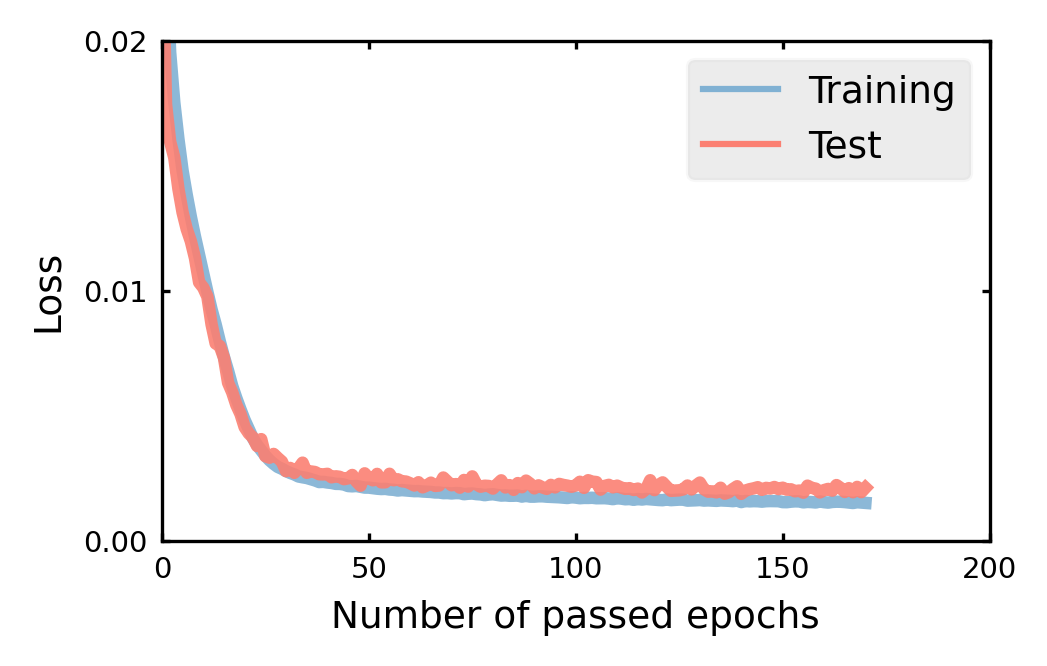}
  \caption{Evolution of the training and test losses during the network training phase. The loss of the test set is closely following the training loss. The small difference between training and test sets at the end of the training phase shows that the network is not over-fitting.}
  \label{fig:training losses}
\end{figure}

When the network architecture is designed, the values of the convolution filter cells and the weights and biases in the dense layers are unknown. During the training phase, these values are "learned" by the neural network. Training means to repeatedly pass a large number of spectra with known labels (training set) through the network and to compare the output of the network with the known input labels of the training set. At the start of the training phase the filter values, weights and biases are initialized randomly. Therefore the predictions of the untrained network will differ strongly from the labels of the input spectra. The difference between the network predictions for the labels and their known values from the input is called "loss" and it is calculated with a loss-function. The loss-function calculates the overall difference between input and output values across all labels. Therefore the loss is a measure of the overall accuracy of the network predictions. An optimisation algorithm is used to slightly change the weights and biases in the network in such a way that, when the training sample is passed through the network again, the loss will be slightly smaller than in the first iteration. Over the course of many iterations of passing the training spectra through the network, calculating the loss and updating the weights and biases for optimisation, the loss steadily decreases and the network predictions get more precise (Figs.~\ref{fig:kiel training} and \ref{fig:training losses}). In the following subsections important concepts that are involved in the training of our neural network are explained in detail.

\subsubsection{Training set and test set}
\label{Training set and test set}

Training relies on a large number of stellar spectra (several thousand) with associated stellar labels. In our case, the labels are previously determined stellar atmospheric parameters and chemical abundances (see Sect.~\ref{Data}). The available data are split randomly into a training set, that is used to train the network, and a test set. During training, the test set is passed through the network as often as the training set, but it is not used in the optimisation calculations that update the weights and biases. Instead, the test set is used to monitor the performance of the network on data that it was not trained on. The loss calculated from the label predictions for the test set is used to determine when to stop the training: If the test loss does not increase anymore over a specified number of training iterations, the weights and biases are assumed to have reached their optimal values for the given network architecture and the training ends. Comparing the performance of the network on the training and test sets also helps to determine if the network is over-fitting. We found that assigning 40\% of our available data to the test set yields the best training results for our application. That means that of our 14\,696 spectra, 8817 spectra are assigned to the training set and 5879 to the test set. Training and test spectra are chosen randomly before the training and it is assured that their labels cover the same parameter space.

\subsubsection{Epochs and batches}
\label{Epochs and batches}

One iteration of passing the entire training set through the network is called an epoch. The number of epochs that are necessary to train a network to achieve good results depends on the model architecture. \par
In one epoch the training data that is passed through the network is divided into equally sized batches. For example, for a training set of 6400 spectra and a batch size of 64, 100 batches pass through the network in one epoch. After every batch that passes through the network, the weights and biases are updated based on the current loss-function in an attempt to decrease the loss after the next batch. This means that in the above example the weights and biases are updated 100 times before the training set has fully passed through the network. This speeds up the overall training because less computer memory is required to process the smaller number of spectra for one update. Using batches can also help to prevent over-fitting. The training spectra are shuffled and assigned to new batches after every epoch.

\section{Training results}
\label{Training results}

We performed ten training runs which resulted in ten slightly different CNN models. The results of the training runs vary slightly because the weights and biases of the network are initialized randomly before every run (the network architecture remains the same). On average one training run lasted for 159 epochs and took $\sim$40 minutes to complete. We removed the two CNN models with the largest remaining test losses at the end of their training phase. The remaining eight CNN models were then used to predict the labels for the spectra in the training, test, and observed sets. The label prediction was very fast: the parameterization of the $\sim$30\,000 spectra in our data set took less than 20 seconds per CNN model. \par
\begin{figure*}
  \centering
  \includegraphics[width=1.6\columnwidth]{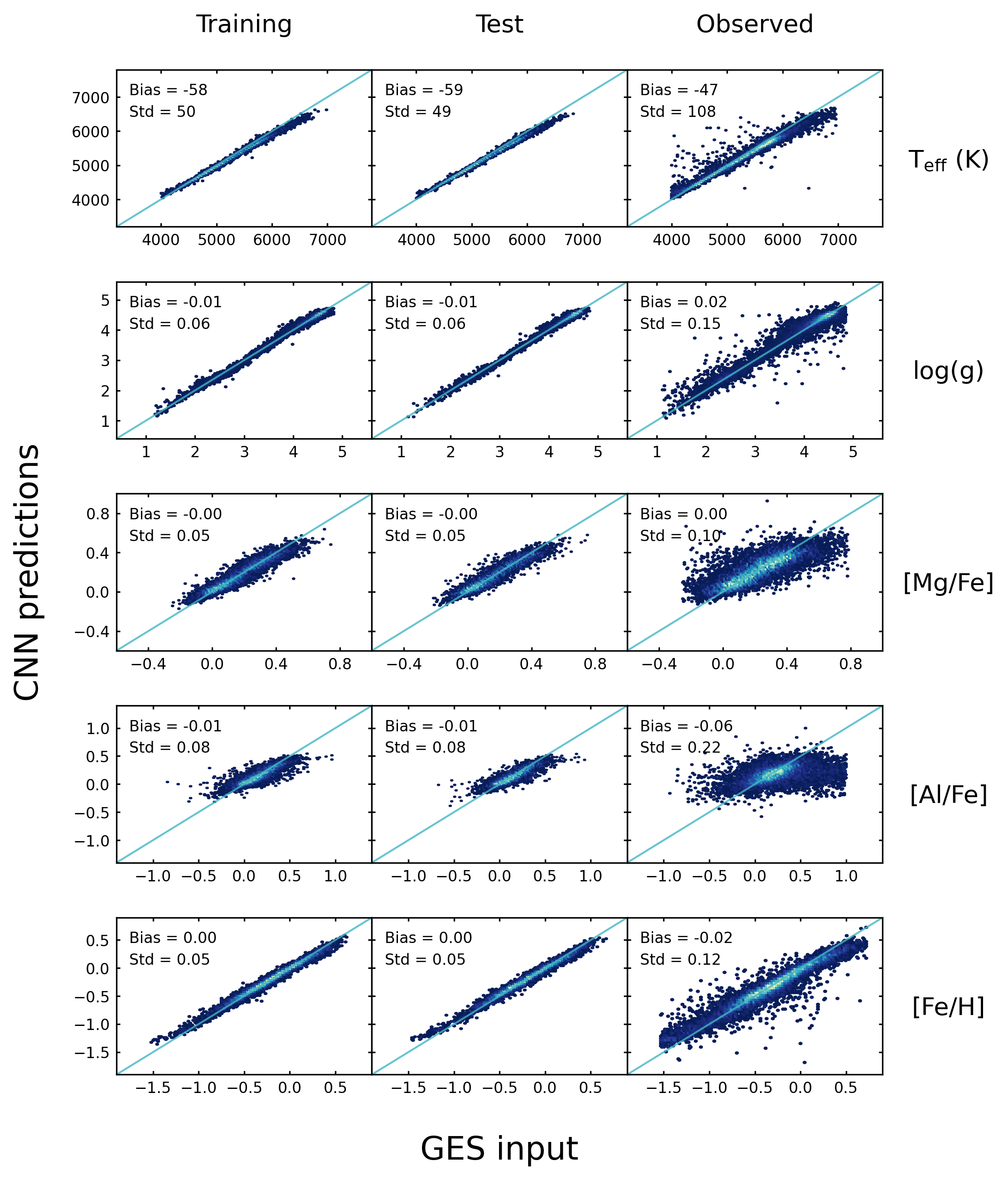}
  \caption{One-to-one comparison of labels from GES iDR6 and the values predicted by our CNN. The three columns show results for the three different data sets (training, test, and observed). Each row contains the results for a different label. In every panel the horizontal axis stands for the GES input labels, the vertical axis represents the labels predicted by our CNN. The average bias and the standard deviation (scatter) of the results around the 1:1 relation are given in every panel. Solid diagonal lines indicate the 1:1 relation.}
  \label{fig:one-to-one}
\end{figure*}

The averages of the eight sets of labels is what we report here as our results. In Fig.~\ref{fig:one-to-one}, we show a direct comparison between the input GES measurements and the CNN predictions for the training, test, and observed sets. There is a good agreement between the GES measurements and CNN predictions across all labels and for all three sets. Both the CNN predictions for the training set and test set show the same offset (if any) and small dispersion around the 1:1 relation. This indicates that the network performs well on spectra which it was not directly trained on and does not overfit. The dispersion around the 1:1 relation is uniform across most of the value ranges of all five labels. However, our CNN does not accurately reproduce the highest and lowest GES measurements. This is especially apparent in the case of [Al/Fe], where the CNN predictions overestimate the lowest [Al/Fe] measurements by $\sim$0.5~dex, while the highest values are underestimated by approximately the same amount. We explain this behaviour by noting that only a small number of spectra with these extreme measurements were available for the network training. The CNN therefore predicts more moderate labels for these spectra. The predictions for the observed set spectra are also in good agreement with the GES input labels, albeit with a larger scatter. The over- and underestimation of the highest and lowest label values is more pronounced in the observed set, especially for the abundance predictions. We expect this poorer performance on the observed set when compared to the training and test sets, because our observed set spectra have a lower S/N between 20 and 30. \par
We further investigated the CNN performance on low quality spectra by constructing a second observed set with S/N between 10 and 20. For this set the predictions for \textit{T}\textsubscript{eff} and log(\textit{g}) are still in good agreement with the GES measurements for most stars. However, the quality of the abundance predictions starts to degrade further, especially for [Mg/Fe] and [Al/Fe]. For these two labels only a weak correlation between GES and CNN values remains. We will show in Sect.~\ref{Learning from spectral features} that our network relies on individual spectral features when predicting abundances. As the quality of the spectra decreases, these features become increasingly hidden by the noise.
 
\subsection{Estimation of internal uncertainties}
\label{Estimation of internal uncertainties}

\begin{figure}
  \includegraphics[width=0.9\columnwidth]{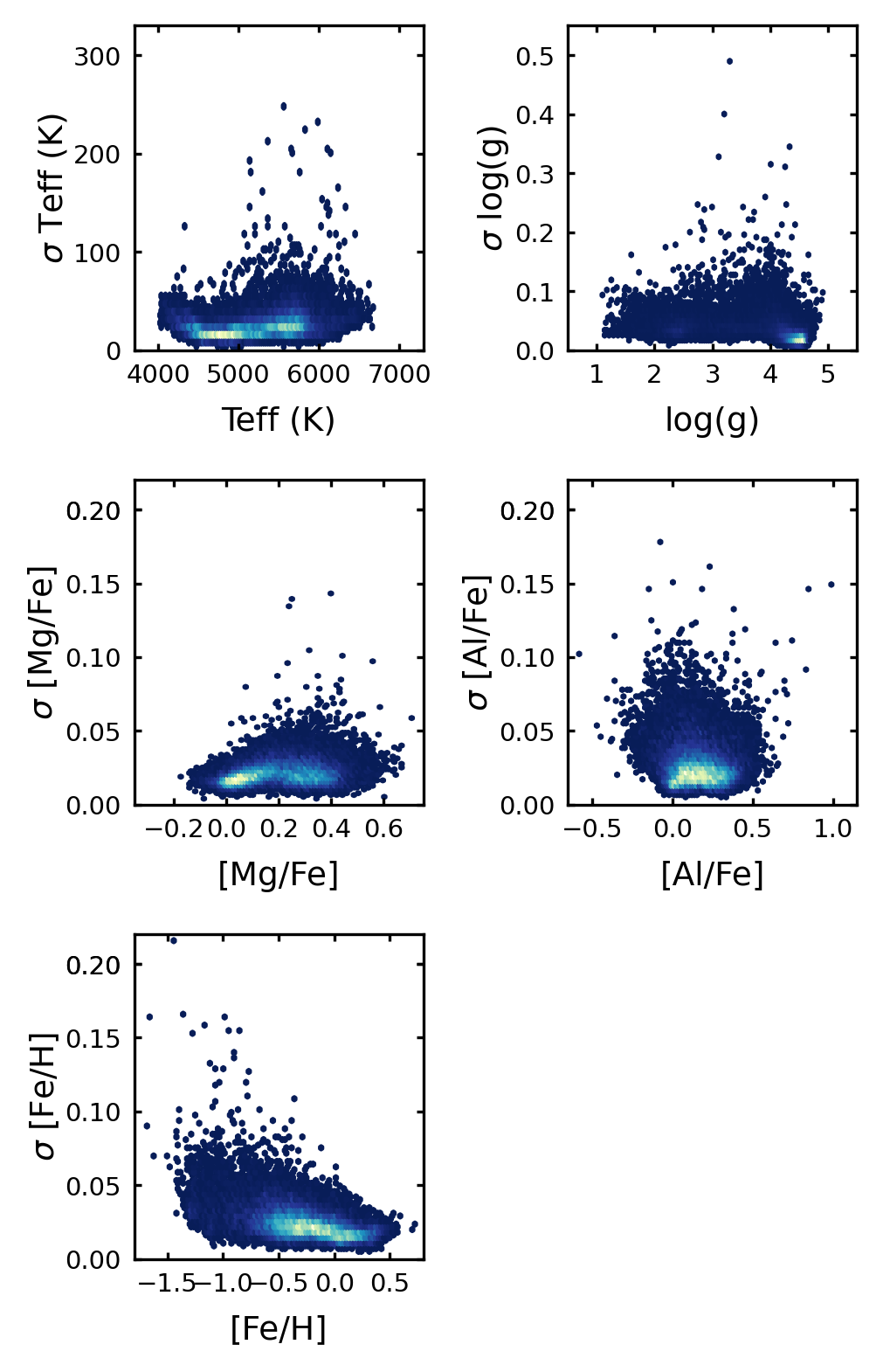}
  \caption{Distributions of label uncertainty as a function of absolute label value for our five parameters. Brighter colors indicate higher density of data points. The displayed data includes all predictions for the training, test, and observed sets.}
  \label{fig:uncertainties}
\end{figure}

As described, the label predictions from our eight trained CNN models vary slightly. This variation can be used to estimate the internal precision of our methodology. We define the uncertainties of our results as the dispersion between the label predictions from the eight CNN models. In Fig.~\ref{fig:uncertainties} we display the distribution of the label uncertainties $\sigma (Label)$ relative to the predicted label values of our five labels. Overall the uncertainties are small with no strong trends with respect to the absolute label values. The exception is [Fe/H], where the errors increase with lower [Fe/H] abundances, presumably due to the smaller number of stars in the metal-poor regime compared to the main bulk of the sample, as well as less spectral features. Also, our CNN struggles to provide precise predictions due to the weak spectral features present in this [Fe/H] regime. The prediction uncertainties for \textit{T}\textsubscript{eff} tend to increase towards the edges of the temperature distribution. The mean uncertainties of the label predictions are small: 24~K for \textit{T}\textsubscript{eff}, 0.03 for log(\textit{g}), 0.02~dex for [Mg/Fe], 0.03~dex for [Al/Fe], and 0.02~dex for [Fe/H]. We find that the uncertainties of the predictions increase as the S/N of the spectra decreases. This is true for spectra from all three sets. 
CNN predictions with large uncertainties for one label also show large uncertainties for all other labels while precise predictions are precise across all five labels. The few outliers with higher uncertainties are results from spectra in the observed set, with lower S/N than the majority of investigated spectra. We further investigate the precision of our CNN in Sect.~\ref{Benchmark stars} by using repeat observations of Benchmark stars.

\subsection{Learning from spectral features}
\label{Learning from spectral features}

The purpose of the convolution layers in our CNN is to find spectral features. These spectral features are then interpreted into the labels by the dense layers. This approach is also used by classical spectral classification methods, where individual spectral features are investigated to derive the stellar parameters. However, since machine--learning is purely data-driven, the predictions of our CNN could merely be the result of our network learning correlations between labels in our data set. Individual elemental abundances for example are correlated with the iron abundance: Stars with low iron generally show low abundances of other elements as well. Inferring stellar parameters from correlations like these can lead to satisfying results for some spectra. However, stars with exotic chemical compositions (for example stars with a non-solar mixture of elements such as old thick disk stars) do not follow such trends and will not be parameterized well. We therefore want to show that our CNN is indeed able to identify spectral lines and to associate them with the right labels. \par
\begin{figure*}
  \centering
  \includegraphics[width=1.6\columnwidth]{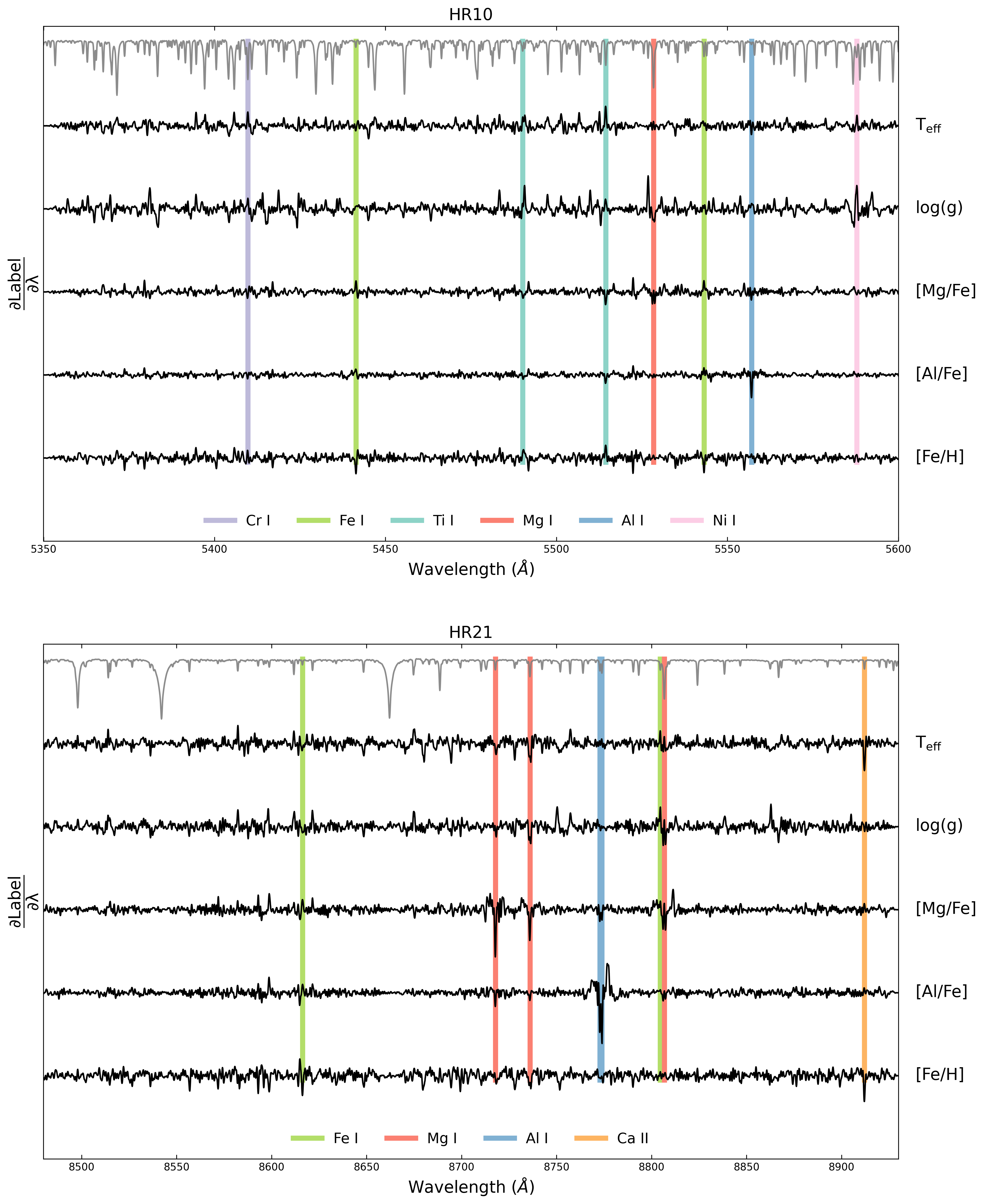}
  \caption{Network gradients for our five labels as a function of wavelength (black). The top panel shows the gradients across the GIRAFFE setup HR10, the bottom panel shows the same for the HR21 setup. An average input spectrum is shown in gray as the top line in both panels. The locations of selected absorption lines of different elements are marked with vertical colored lines. The highlighted Mg and Al lines were used by GES for the determination of our input Mg and Al abundances. Their wavelengths are 5528.41, 8717.81, 8736.02 and 8806.756~$\AA$ for Mg and 5557.06, 8772.87 and 8773.90~$\AA$ for Al \citep{2021A&A...645A.106H}.}
  \label{fig:network gradients}
\end{figure*}

During the training phase the optimisation algorithm calculates the sensitivity of the output labels to small changes in the input flux values. This is done for every wavelength bin in the input spectrum. It is therefore inherent to neural network training to calculate which output label is sensitive to which portion of the input spectrum. The sensitivity of the output labels to the flux at a certain wavelength bin $\lambda$ can be expressed as the gradient $\partial$ Label / $\partial \lambda$. A large absolute gradient value at a wavelength bin then means that the network is very sensitive to flux changes in that bin. In Fig.~\ref{fig:network gradients} we show the network gradients for our five labels across the whole wavelength range of the input spectra. The gradients are scattered randomly around zero for most of the wavelength range. Only at certain wavelength bins the network is sensitive to flux changes. Here, the gradients show individual, narrow spikes. This is especially apparent in the gradients for [Mg/Fe] and [Al/Fe] in the HR21 part of our input spectra. The [Mg/Fe] gradients show two clear spikes at 8736.0 and 8806.8~$\AA$. These are the locations of two \ion{Mg}{i} absorption lines. The largest spike in the [Al/Fe] gradients mark the location of the \ion{Al}{i} double feature at $\sim$8773~$\AA$. We therefore see that our network is able to identify absorption lines in the input spectra. The negative gradient values at these wavelengths means that if the flux at the absorption lines are low, the predicted abundance is high, and vice-versa. This reflects the fact that stronger absorption features in spectra indicate higher elemental abundances in stellar atmospheres. The CNN label predictions are therefore directly based on the strength of the relevant absorption lines in the input spectra. \par
Our network does not only learn from the correlation between spectral features and stellar labels in individual stars, but also from correlations between labels across the whole training set. These data-wide correlations are of astrophysical origin, showing for example that stars with high iron abundance generally also have high abundances of other metals. To investigate how astrophysical correlations in the input data influence the network gradients, we trained our CNN with different combinations of input labels. We found that the gradients of a combination of \textit{T}\textsubscript{eff}, log(\textit{g}), and one or all of the abundances show no gradient correlations, meaning the CNN learns mainly from the spectral features. If the network is trained only with the highly correlated labels A(Mg), A(Al) and A(Fe) (absolute abundances), the gradients for the three labels are almost identical. In this case the CNN is still able to identify the locations of the Mg, Al and Fe absorption lines, but the network predictions for one element is also very sensitive to absorption lines of the other two elements. In addition, the quality of the CNN predictions starts to degrade, leading to larger differences between GES input labels and CNN predictions. This is because the network relied too much on the label correlations within the training set instead of the connection between spectral features and labels of individual spectra. For future surveys, we therefore recommend to carefully inspect the training data for strong correlations because they can influence the CNN performance. \par
Further investigation of the gradient peaks gives interesting insights into the behaviour of our CNN. Some spectral lines influence the network predictions for only one of the labels. An example in the HR10 setup is a \ion{Cr}{i} line at $\sim$5410 $\AA$, that corresponds to a peak in the gradient for \textit{T}\textsubscript{eff}. Other lines have an effect on multiple, uncorrelated labels. For deriving \textit{T}\textsubscript{eff} and log(\textit{g}), our CNN is sensitive to the \ion{Ni}{i} line in the red end of the HR10 setup. While this line coincides with the strongest peak in the log(\textit{g}) gradient, only a minor peak is present in the \textit{T}\textsubscript{eff} gradient. A \ion{Fe}{i} line at $\sim$8805~$\AA$ is also important for both the \textit{T}\textsubscript{eff} and log(\textit{g}) predictions, but not for the [Fe/H] likely due to its blend with a Mg line.
The infrared calcium triplet (the three most prominent absorption lines in the HR21 setup) does not have a significant influence on the network predictions for any of the labels, but the \ion{Ca}{ii} line beyond 8900~$\AA$ causes a very strong response of the \textit{T}\textsubscript{eff} and [Fe/H] gradients. A deeper investigation of the CNN gradients could be done to search for complementary spectral features that could be used by standard spectroscopic pipelines, but this is out of the scope of the present paper.

\section{Validation of results}
\label{Validation of results}

In this section we validate our results in two ways. First, we compare CNN results to the GES input labels for a set of benchmark stars. In this way we can validate that our CNN can accurately parameterize individual spectra. Then we investigate the label predictions for spectra from stars in different stellar populations to confirm that our results recover important Milky Way properties. Our validation covers the results from our whole sample of spectra, combining training, validation, and observed sets. 

\subsection{Benchmark stars}
\label{Benchmark stars}

\begin{figure}
  \centering
  \includegraphics[width=0.9\columnwidth]{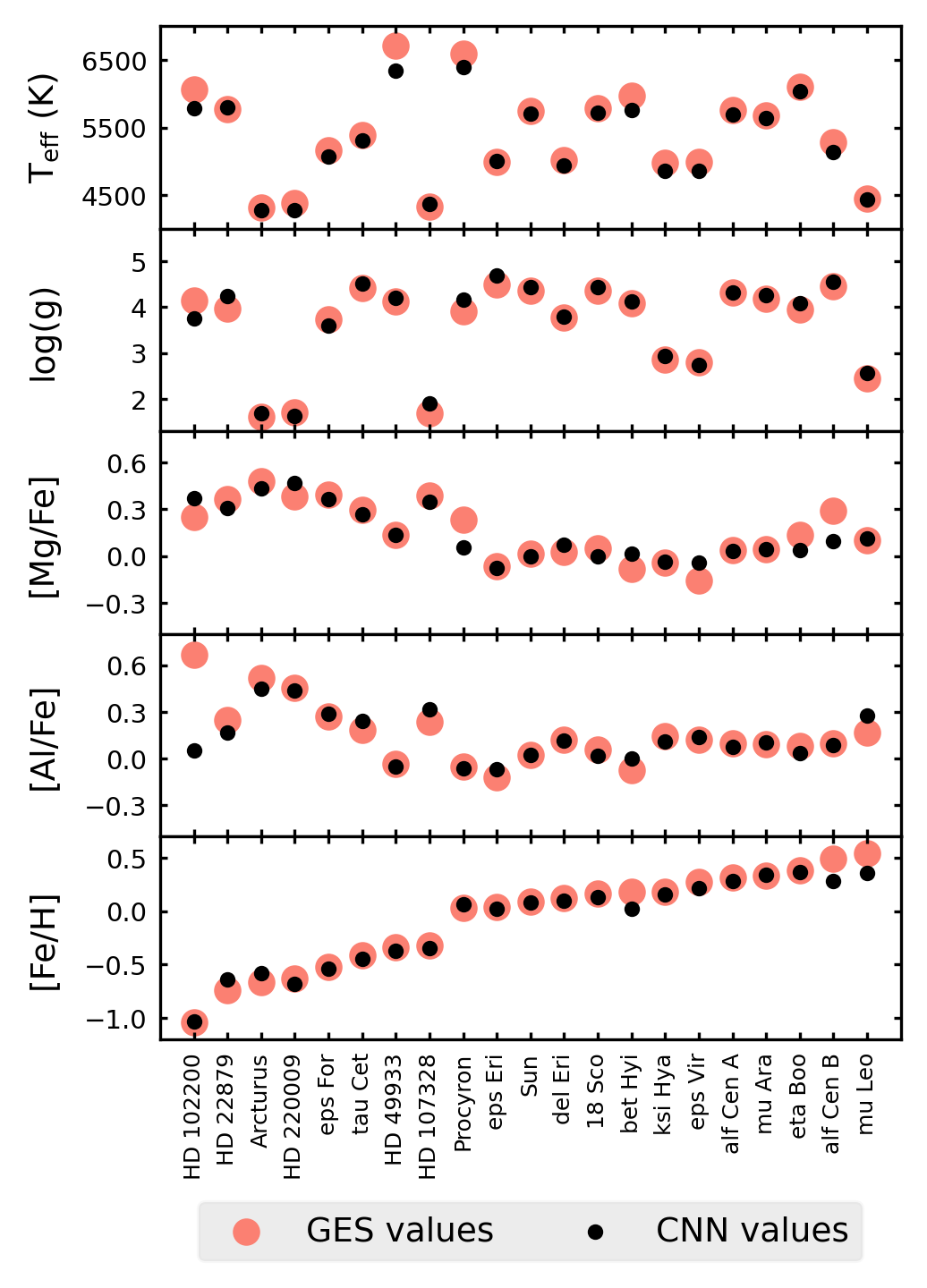}
  \caption{Comparison of GES input labels with CNN predictions for the benchmark stars. Different data point sizes have no physical meaning and are for visualisation purposes only.}
  \label{fig:benchmarks}
\end{figure}

The GES iDR6 data set contains a number of benchmark stars with high quality spectra and precise stellar labels \citep{2015A&A...582A..49H}. This benchmark set covers stars in different evolutionary stages with a wide range of stellar parameters and abundances, suited for the verification and calibration of large data sets \citep{2017A&A...598A...5P}. Our data set contains 25 benchmark stars, including the Sun. For our analysis we excluded four benchmark stars that have labels outside our training label space. Three of the excluded stars have [Fe/H] < $-2.0$~dex and one star has a [Mg/Fe] abundance of $-0.79~dex$. We compare the GES labels and CNN output labels for the remaining 21 benchmark stars in Fig.~\ref{fig:benchmarks}. The CNN predictions agree well with the GES values across all five labels for most of the benchmark stars. The largest differences occur for stars on the edges of the parameter space, where the network predicts more moderate values compared to the extreme GES values. An example is HD 49933, the benchmark star with the highest \textit{T}\textsubscript{eff}, for which our network predicts $\sim$350~K less than what is reported by GES. This star remains one of the hottest in our benchmark set, even with this reduction in \textit{T}\textsubscript{eff}. Despite the large difference in one label, the CNN predictions for the other labels of HD 49933 agree well with the GES measurements. The only star that shows a large difference across several labels is HD 102200. It has the highest GES [Al/Fe] measurement of the benchmark stars, which our network underestimates by $\sim$0.6~dex. \par
The label-specific bias and scatter between GES and CNN labels for the benchmark stars is comparable to the bias and scatter that we found for the training and test sets in Fig.~\ref{fig:one-to-one}. \par
The CNN predicts similar label values for repeat spectra, oftentimes predicting identical labels for multiple repeats. The dispersions between repeated label predictions can be interpreted as the uncertainties of the CNN results. These CNN uncertainties are within the GES label uncertainties for the benchmark stars. \par
We conclude that our CNN is able to accurately predict multiple labels of individual stars. However, the most extreme CNN results should be used cautiously, because they are likely underestimating high values and overestimating low values.

\subsection{Comparison to asteroseismic surface gravities}
\label{Comparison of CNN log(g)s with asteroseismic surface gravities}

\begin{figure}
  \centering
  \includegraphics[width=0.9\columnwidth]{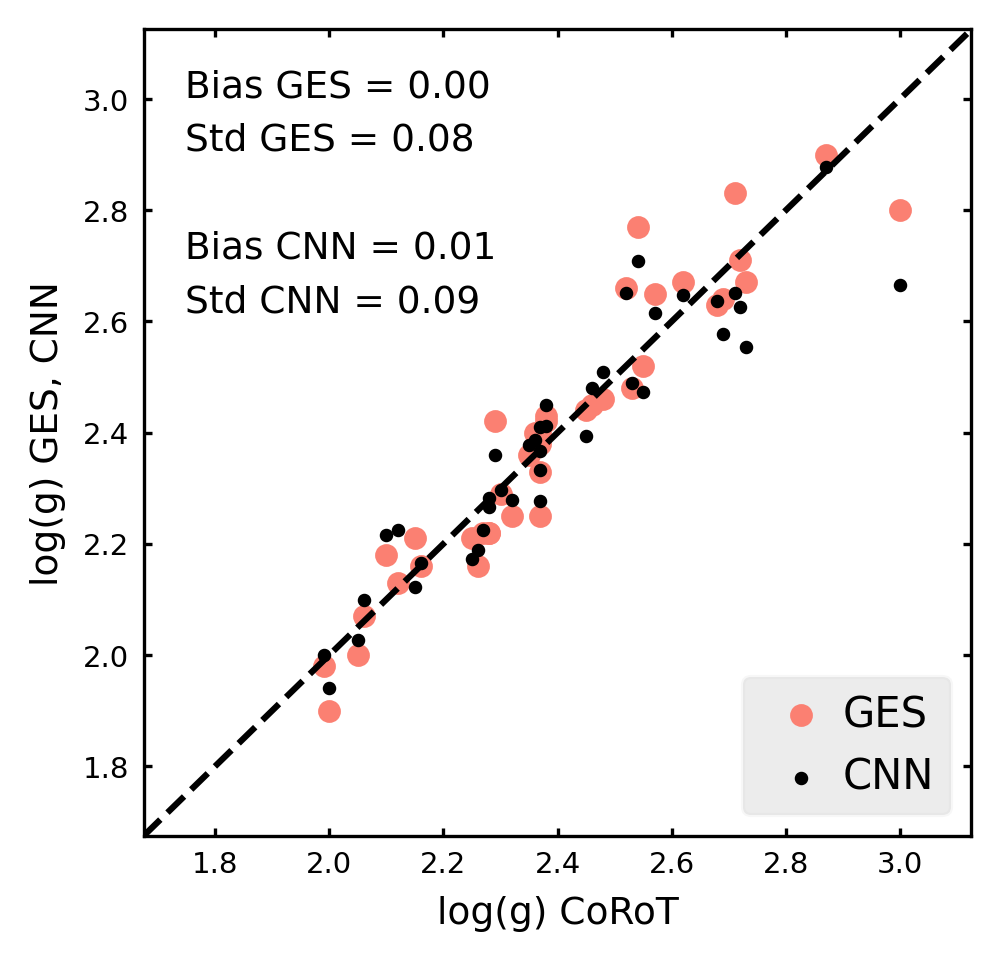}
  \caption{One-to-one comparison between input labels from GES in red (output CNN label in black) with respect to seismic log(\textit{g}) derived using asteroseismology \citep{valentini2016}.}
  \label{fig:1v1_CoRoT}
\end{figure}

Asteroseismology is an extremely powerful tool to provide accurate surface gravities, based on stellar oscillations. This method is massively used by spectroscopic surveys for validation or calibration purposes (RAVE, \citealt{Valentini2017}; APOGEE, \citealt{Pinsonneault2018}). The Convection, Rotation and planetary Transits (CoRoT) mission was a space observatory dedicated to stellar seismology.
Our aim here is to compare the log(\textit{g}) values of our GES input data and our CNN results to GES-CoRoT log(\textit{g}) values from \cite{valentini2016}. In Figure \ref{fig:1v1_CoRoT}, the comparison between GES log(\textit{g}) values and asteroseismic CoRoT results shows no residual trend, with a low dispersion of 0.08~dex. The CNN log(\textit{g}) values show also no residual trend compared to GES-CoRoT log(\textit{g}) and a similarly small dispersion.

\subsection{Globular clusters}
\label{Globular clusters}

\begin{figure}
  \includegraphics[width=0.9\columnwidth]{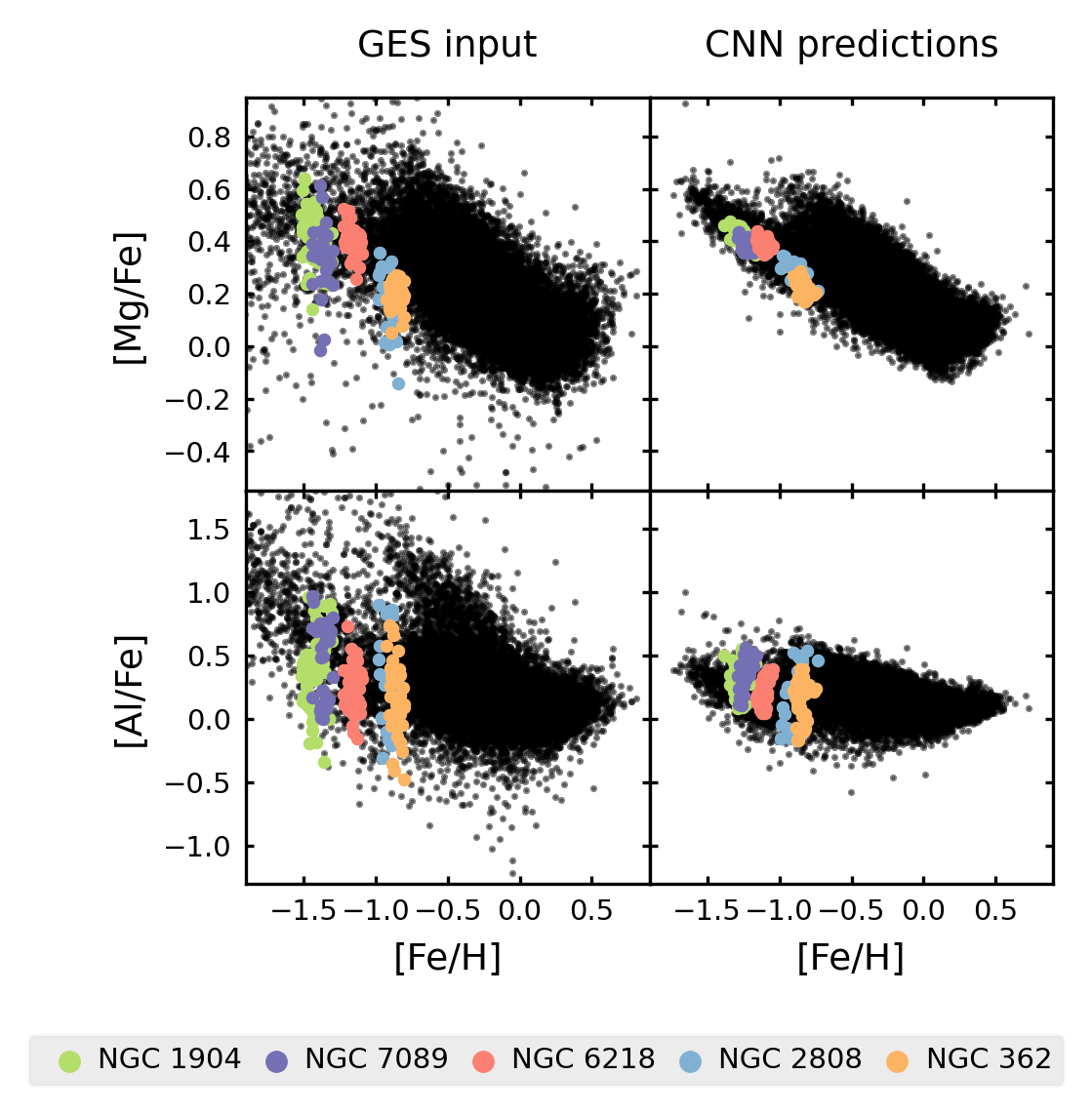}
  \caption{[Mg/Fe] and [Al/Fe] vs. [Fe/H] plots for stars in the training, test, and observed sets. The panels on the left show the distributions of the GES iDR6 values, panels on the right are the predictions of the trained neural network. Cluster membership is indicated by differently colored data-points.}
  \label{fig:clusters}
\end{figure}

Our data set covers stars that belong to a number of different globular clusters. We identified member stars of five separate clusters based on their position in the sky and their scatter in [Fe/H] and radial velocities that are reported in GES iDR6. The position of the cluster members in the [Mg/Fe] and [Al/Fe] plots is displayed in Fig.~\ref{fig:clusters}. The CNN predictions reproduce the grouping of cluster members in the plots, with a small spread of [Fe/H] within each cluster. However, the CNN predictions show a smaller scatter in [Element/Fe] compared to the GES values, especially for Al. This reduced scatter is a reflection of the results that we saw in Fig.~\ref{fig:one-to-one}, where the CNN predicts more moderate labels for spectra with extreme GES labels.

\begin{figure}
  \centering
  \includegraphics[width=0.9\columnwidth]{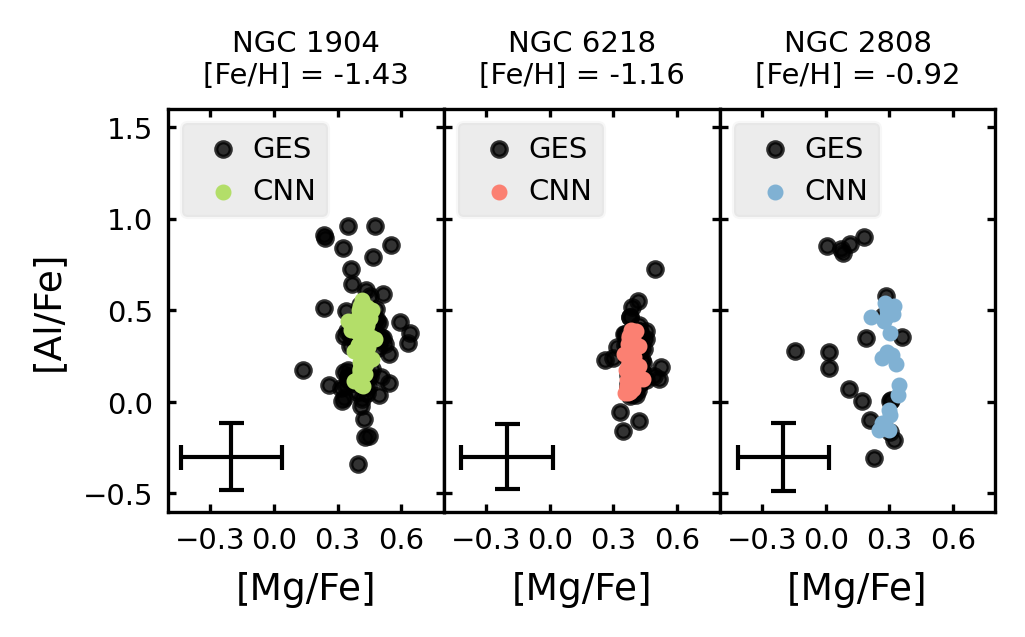}
  \caption{Mg-Al anti-correlation plots for three of our sample clusters with decreasing cluster metallicity. Colored data points show the labels predicted by the CNN, black points are the GES results. The color for the differnet clusters is the same as in Fig.~\ref{fig:clusters}. Average uncertainties of the GES results are shown in the lower left corner.}
  \label{fig:MgAl_3clusters}
\end{figure}

Our CNN results recover the Mg-Al anti-correlation, which is used to investigate the chemical evolution of globular clusters \citep{2017A&A...601A.112P}. Figure \ref{fig:MgAl_3clusters} shows the Mg-Al anti-correlation in the clusters NGC 1904, NGC 6218 and NGC 2808. The average [Fe/H] values of these three clusters span a range of $\sim$0.5~dex. We see that the CNN results trace well the anti-correlations in all three clusters. Except for the most extreme stars, all CNN predictions agree with the GES results within their reported uncertainties. We observe the largest difference between CNN and GES labels in NGC 1904, which contains stars with the lowest [Fe/H] in our entire data set.

\subsection{Thin- and thick disk populations}
\label{Thin- and thick disk populations}

\begin{figure}
  \centering
  \includegraphics[width=0.9\columnwidth]{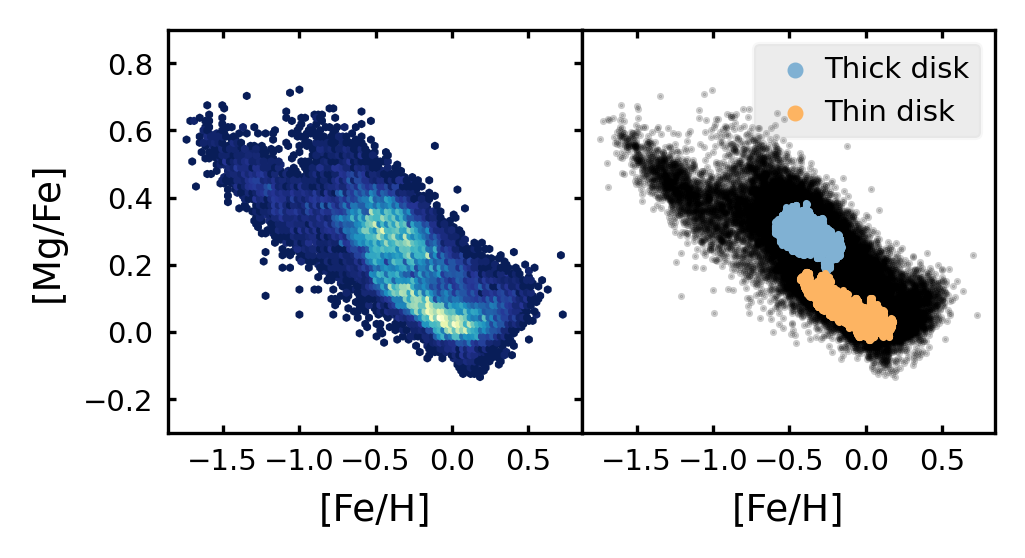}
  \includegraphics[width=0.9\columnwidth]{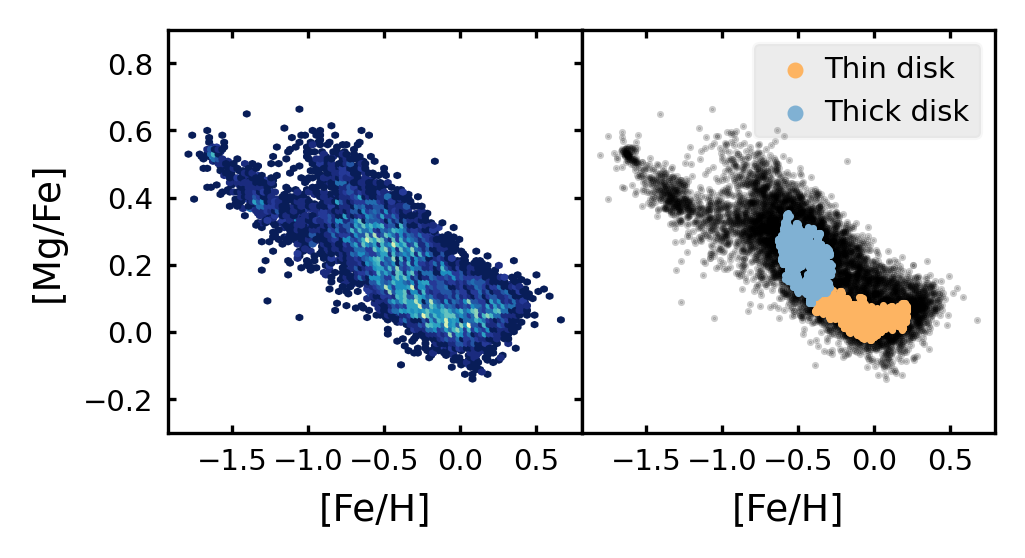}
  \caption{Top panel: Density map of the [Mg/Fe] vs. [Fe/H] distribution of our CNN results for the training+test sample. Brighter colors indicate a higher density of data points. Thin and thick disk populations found by the HDBSCAN algorithm are shown on the right. The two populations correspond to the two separate dense regions on the left panel. Bottom: same plot but for the observed sample composed of 15\,419 stars with 20 $\le$ S/N $\le$ 30.}
  \label{fig:disks}
\end{figure}

As discussed in Sect.~\ref{Parameter space of input labels}, [Mg/Fe] values can be used to separate the Milky Way stars into a thin disk and a thick disk population. We performed this separation based on our CNN results for the combination of training plus test set and and, independently, for the observed set. The top panel of Fig.~\ref{fig:disks} shows the distribution of [Mg/Fe] vs. [Fe/H] for the CNN predictions (analogous to the top panel of Fig.~\ref{fig:MgAl_Fe_hist2d}) for the training/test samples. We can see the separation between the two main populations: Thin disk stars with [Mg/Fe] lower than $\sim$0.2~dex and thick disk stars with enhanced [Mg/Fe]. To identify thick and thin disk stars in our data set, we used the clustering algorithm HDBSCAN \citep{10.1007/978-3-642-37456-2_14}, which is implemented in the \textit{hdbscan} library for Python programming. This algorithm assigns data points to different clusters, depending on the density of data points in a distribution. The result of this clustering for our CNN data is displayed in the right panel of Fig.~\ref{fig:disks}. Two clusters are identified that correspond to the two stellar populations. About 35\% of the stars do not fall into any of the two clusters. Stars outside of the two dense regions in the distribution are considered to be "noise" by the HDBSCAN algorithm and are not assigned to any cluster. In the literature the chemical separation between thin and thick disk is often performed by splitting the distribution into several [Fe/H] bins and finding the [Mg/Fe] value in each bin where the density of stars is at a minimum (e.g. \citealt{2011A&A...535L..11A}, \citealt{2014A&A...572A..33M}). \cite{2018A&A...619A.125A} use a sophisticated t-SNE approach to identify the different stellar populations. They include abundances measurements from 13 chemical elements to further dissect the thin and thick disk into additional sub-populations. \par
We proceeded the same investigation for the observed sample (covering 20 $\le$ S/N $\le$ 30) to test the precision of CNN abundances in a regime of low S/N. The results are shown in the bottom panels of Fig.~\ref{fig:disks}. The HDBSCAN algorithm is able to identify the same two clumps corresponding to the thin and the thick discs but, instead of two separated blobs, they seem to form a single sequence from low to high [Mg/Fe]. Machine--learning on low-S/N spectra may not be able to derive abundances precisely enough for Galactic Archaeology. We therefore warn the community that spectra with high-enough S/N should be gathered by surveys in order to perform Galactic Archaeology.

To investigate the age distributions of the two populations, we used the isochrone fitting code \textit{A Unified tool to estimate Distances, Ages and Masses} (UniDAM). The UniDAM tool \citep{2017A&A...604A.108M} follows a Bayesian approach of isochrone fitting. It compares stellar atmospheric parameters and absolute magnitudes from simulated PARSEC isochrones \citep{2012MNRAS.427..127B} to the corresponding values in observed stars. All PARSEC isochrones also have stellar masses and ages associated to them. For the isochrone fitting we used the CNN predictions for the atmospheric parameters \textit{T}\textsubscript{eff} and log(\textit{g}) in combination with [Fe/H]. Magnitudes of our sample stars in the \textit{J}, \textit{H}, and \textit{K} bands were taken from the 2MASS catalogue \citep{2006AJ....131.1163S}. In order for UniDAM to calculate the absolute magnitudes, it is also necessary to provide the parallax value for each sample star. We used the parallaxes from Gaia EDR3 \citep{2020yCat.1350....0G}. We removed stars with negative parallaxes as well as stars with relative parallax errors > 0.2. To get the most precise age estimates, we only considered turn-off stars in this analysis. Turn-off stars in our thin and thick disk samples were selected by their position in the Kiel-diagram. The resulting average age of the thin disk stars is 8.8~Gyr, the average thick disk age is 9.8~Gyr. This age difference between the two populations has been found in numerous studies and by using several different age determination methods. \cite{Kilic_2017} for example find ages from 7.4–8.2~Gyr for the thin disk and 9.5–9.9~Gyr for the thick disk by analyzing luminosity functions of white dwarfs in the two disks. Using APOGEE spectra and precise age estimates based on asteroseismic constraints, \cite{2021A&A...645A..85M} also show that the chemically selected thick disk stars are old, with a mean age of $\sim$11~Gyr. We note that the detailed age distribution of thin and thick disk members is sensitive to several selection criteria such as metallicity, kinematic properties and the distance from the Milky Way center. A detailed investigation of the two stellar populations is out of the scope of this work. \par
\begin{figure}
  \centering
  \includegraphics[width=0.9\columnwidth]{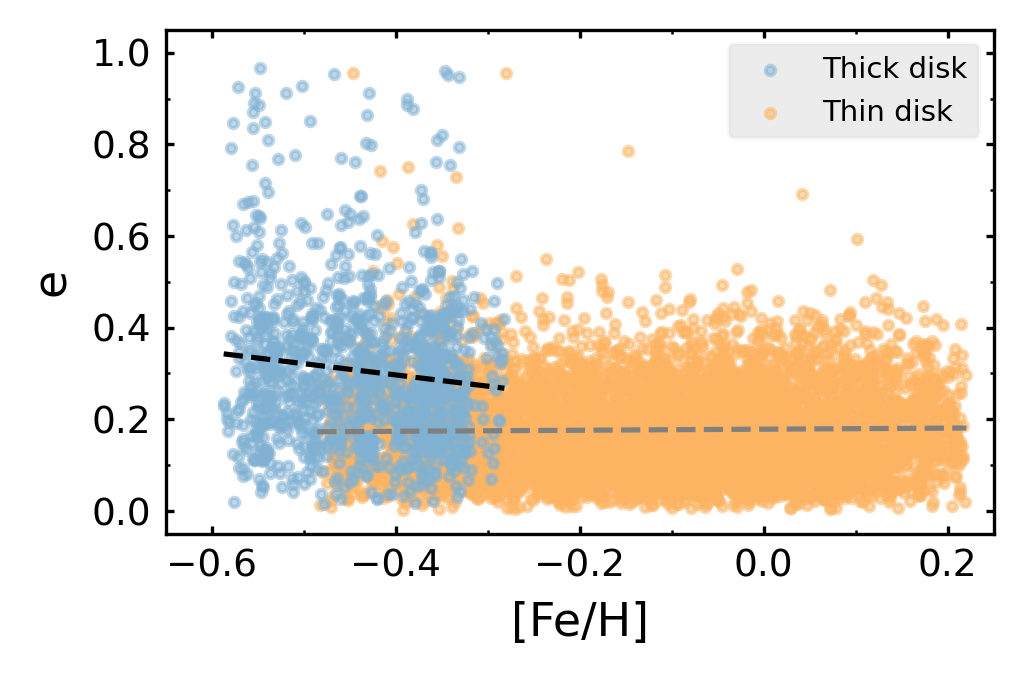}
  \caption{Eccentricities $e$ of stellar orbits as a function of [Fe/H] for our thin disk and thick disk samples. Dashed lines show linear fits to the thick disk (black) and thin disk data points (gray).}
  \label{fig:eccentricities}
\end{figure}

We also investigated the kinematical properties of our thin and thick disk samples. Based on the current positions and velocities of the stars, we integrated their orbits for 5 Gyr in a theoretical Milky Way potential, using the python-based galactic dynamics package \textit{galpy} \citep{2015ApJS..216...29B}. For the integration we used the gravitational potential \textit{MWPotential2014}, which combines bulge, disk and halo potentials. Proper motions, sky coordinates and parallaxes of our sample were taken from the Gaia EDR3. In Fig.~\ref{fig:eccentricities} we show the trends of the orbital eccentricities relative to [Fe/H] for our thick and thin disk stars. A linear regression model shows that the eccentricity $e$ of thick disk orbits is decreasing with increasing [Fe/H]: $\Delta e / \Delta\rm[Fe/H]$ = $-0.25$. The eccentricities of thin disk stars are on average lower than the thick disk eccentricities and show a slight positive trend ($\Delta e / \Delta\rm[Fe/H]$ = 0.01). These results are consistent with the findings of \cite{Yan_2019}, who investigated the chemical and kinematical properties of thin and thick disk stars from the LAMOST data set \citep{2012RAA....12..723Z}.

\section{Caveats}
\label{Caveats}

During the network training the GES input labels are considered to provide the true parameterization of the training spectra. The quality of the network predictions therefore depends entirely on the quality of the training data. We limited the uncertainties and errors in our training data by applying several quality constraints (Sect.~\ref{Data preparation}), but there is still a possibility that the input labels may suffer from systematics. Inaccurate labels of a small number of input spectra will not have a noticeable effect on the training process. 
The cases with a large difference between GES input value and CNN prediction could therefore be the result of the network predicting accurate labels for spectra with inaccurate GES labels. Future work could investigate if and how CNNs can be used for the quality control of classically derived stellar parameters. \par 
We are able to estimate the internal uncertainties of our network predictions by training multiple CNN models on the same data. These uncertainties however do not take into account the uncertainties of the training labels themselves. Bayesian deep learning frameworks account for both the training data uncertainties and model uncertainties \citep{10.5555/3295222.3295309}. Future work could benefit from implementing this Bayesian approach into our CNN method. \par
The predictive power of our CNN is limited by the sparse training data that is available at the edges of the parameter space (Sect.~\ref{Training results}). A more homogeneous coverage of the parameter space, achieved by increasing the number of training spectra with extreme parameter values, will increase the precision of the CNN predictions. \par
During the training phase our CNN not only learns the correlations between spectral features and stellar labels, but is also sensitive to correlations within the training labels themselves. The effect of this is discussed in Sect.~\ref{Learning from spectral features}, where we see for example how the strength of Mg absorption lines also has an effect on the network predictions for [Al/Fe]. These correlations can never be avoided when training the network to predict multiple abundances at once. The alternative then is to train a separate network model for each abundance label. This strategy decreases the efficiency of the CNN approach, especially when the goal is to predict a large number of chemical elements. Therefore, care has to be taken to reduce the correlations in the training data without sacrificing the ability of the network to predict multiple labels at once. \par

\section{Conclusions}
\label{Conclusions}

Here we summarize the main results of our study and the steps we carried out to find these results.

\begin{itemize}
    \item
    We built a training and a test set based on GES iDR6 spectra with S/N > 30. Together, these sets consist of 14\,696 stellar spectra with associated atmospheric parameters and chemical abundances. We applied several quality checks on these sets to ensure that our network is trained on high quality spectra and stellar labels. We use the parameters \textit{T}\textsubscript{eff} and log(\textit{g}) and the abundances [Mg/Fe], [Al/Fe], and [Fe/H] as the input labels for our neural network. We also built an observed set of 15\,419 spectra to test the performance of our CNN on spectra that were not involved in the training process. The observed set spectra have a lower S/N between 20 and 30.
    \item
    We then built a convolutional neural network with the python-based library \textit{Keras}. Our network architecture contains three convolutional layers, designed to detect features and absorption lines in input spectra. Three succeeding dense layers then convert the found spectral features into the values of the five output labels. We performed ten training runs, resulting in ten slightly different CNN models. We used the eight best CNN models to predict the labels of the training, test, and observed set spectra.
    \item
    The CNN label predictions are in good agreement with the GES input labels. The bias (average offset) and scatter between CNN and GES labels are identical for the training and test sets, showing that our CNN is not over-fitting during the training. The results for the observed set are also in good agreement with the GES input values, albeit with a larger scatter between CNN and GES values. We find that the quality of the CNN results degrades for low S/N spectra (20 $\le$ S/N $\le$ 30), especially for abundance predictions. We warn the community that machine--learning on low-S/N spectra may not be sufficient for deriving precise enough abundances. Surveys should therefore gather spectra with high-enough S/N (depending on their science goals).
    \item
    All three sets have in common that the differences between CNN predictions and GES values increase at the edges of the parameter space. Here the number of available training spectra is small. Increasing the number of training spectra in these parameter regimes can increase the precision of the CNN predictions.
    \item
    The scatter between the predictions from the eight different CNN models can be used to assess the internal precision of our network. This scatter is small: 24~K for \textit{T}\textsubscript{eff}, 0.03 for
    log(\textit{g}), 0.02~dex for [Mg/Fe], 0.03~dex for [Al/Fe], and 0.02~dex for [Fe/H].
    \item
    We use network gradients to demonstrate the sensitivity of our network to different parts of the input spectra. The gradients show that the network is able to identify absorption lines in the input spectra and associates those lines to the relevant stellar labels. Caution has to be applied when choosing input labels, because strongly correlated input labels lead to strongly correlated network gradients. The network then predicts labels based on unrelated spectral features (for example absolute Al abundance from Mg absorption lines). Inferring stellar parameters from correlations like these can lead to satisfying results for some spectra. However, stars with exotic chemical compositions will not be parameterized well.
    \item
    The validation of our results with 21 GES benchmark stars shows that our CNN is able to precisely predict labels for individual stars over a large range of label values. Network predictions for repeat spectra of the benchmark stars show a small scatter per star. This scatter is within the GES uncertainties for the benchmark star labels.
    \item
    We investigated the Mg-Al anti-correlation in globular clusters, ranging from -0.92 to -1.5 in metallicity.
    As our training set do not contain a large number of stars in the metal-poor regime with large Al and Mg abundances, CNN mainly recovered the spread in Al in GCs.
    \item
    We investigated the ages and chemical properties of the galactic thin and thick disk populations. We identified thin- and thick disk stars based on their position in the [Mg/Fe] vs. [Fe/H] plane with the HBDSCAN algorithm. We find the average age of the thin disk stars to be 8.8~Gyr and 9.8~Gyr for the thick disk stars. The orbit eccentricities of the thick disk stars show a negative trend with [Fe/H] ($\Delta e$/$\Delta$[Fe/H] = $-0.25$). The eccentricities of thin disk orbits are lower than those of the thick disk and show no significant trend with [Fe/H]. These results, based on our CNN predictions, are consistent with similar results in the literature.
\end{itemize}

Our study is of a significant importance for the exploitation of future large spectroscopic surveys, such as WEAVE and 4MOST. We showed that CNN is a robust methodology for stellar parametrization, while we raised some caveats that should be taken into by the community for future use of ML algorithms in general.

\begin{acknowledgements}

These data products have been processed by the Cambridge Astronomy Survey Unit (CASU) at the Institute of Astronomy, University of Cambridge, and by the FLAMES/UVES reduction team at INAF/Osservatorio Astrofisico di Arcetri. These data have been obtained from the Gaia-ESO Survey Data Archive, prepared and hosted by the Wide Field Astronomy Unit, Institute for Astronomy, University of Edinburgh, which is funded by the UK Science and Technology Facilities Council.
This work was partly supported by the European Union FP7 programme through ERC grant number 320360 and by the Leverhulme Trust through grant RPG-2012-541. We acknowledge the support from INAF and Ministero dell' Istruzione, dell' Universit\`a' e della Ricerca (MIUR) in the form of the grant "Premiale VLT 2012". The results presented here benefit from discussions held during the Gaia-ESO workshops and conferences supported by the ESF (European Science Foundation) through the GREAT Research Network Programme.)
This work has made use of data from the European Space Agency (ESA) mission
{\it Gaia} (\url{https://www.cosmos.esa.int/gaia}), processed by the {\it Gaia}
Data Processing and Analysis Consortium (DPAC,
\url{https://www.cosmos.esa.int/web/gaia/dpac/consortium}). Funding for the DPAC
has been provided by national institutions, in particular the institutions
participating in the {\it Gaia} Multilateral Agreement.
This publication makes use of data products from the Two Micron All Sky Survey, which is a joint project of the University of Massachusetts and the Infrared Processing and Analysis Center/California Institute of Technology, funded by the National Aeronautics and Space Administration and the National Science Foundation.
This article is based upon work from COST Action CA16117, supported by COST (European Cooperation in Science and Technology).
T.B. was supported by grant No. 2018-04857 from the Swedish Research Council.
M.B. is supported through the Lise Meitner grant from the Max Planck Society. We acknowledge support by the Collaborative Research centre SFB 881 (projects A5, A10), Heidelberg University, of the Deutsche Forschungsgemeinschaft (DFG, German Research Foundation).  This project has received funding from the European Research Council (ERC) under the European Union's Horizon 2020 research and innovation programme (Grant agreement No. 949173)

\end{acknowledgements}

\bibliography{references}
\bibliographystyle{aa}

\end{document}